# Role of contact bonding on electronic transport in metal-carbon nanotube-metal systems

**Ioannis Deretzis and Antonino La Magna**

Istituto per la Microelettronica e Microsistemi (CNR-IMM), Stradale Primosole 50, 95121 Catania, Italy.

E-mail: ioannis.deretzis@imm.cnr.it, antonino.lamagna@imm.cnr.it

**Abstract:** We have investigated the effects of the interfacial bond arrangement on the electronic transport features of metal-nanotube-metal systems. The transport properties of finite, defect-free armchair and zigzag single-walled carbon nanotubes attached to Au(111) metallic contacts have been calculated by means of the non-equilibrium Green functional formalism with the Tight-Binding and the Extended Hückel Hamiltonians. Our calculations show that the electrode material is not the only factor which rules contact transparency. Indeed, for the same electrode, but changing nanotube helicities, we have observed an overall complex behaviour of the transmission spectra due to band mixing and interference. The comparison of the two models shows that the Tight Binding approach fails to give a satisfactory representation of the transmission function when a more accurate description of the C-C and Au-C chemical bonds has to be considered. We have furthermore examined the effect of interface geometry variance on conduction and found that contact-nanotube distance has a significant impact, while contact-nanotube symmetry plays a marginal, yet evident role.

## 1. Introduction

The microelectronics industry's interest for a potential use of carbon nanotubes (CNTs) as building blocks of nanoscale devices is well founded. As soon as the particularity of the electrical and mechanical properties of CNTs became known (diameter and folding angle dependent conducting behaviour, mechanical strength, thermal conductivity, elasticity, manipulation ability), extensive studies focused on the argument and prototype devices (e.g. field-effect transistors [1-3]) were designed and implemented with the objective of a future substitution of current components of digital electronic circuits with new molecular ones. The role of the metal contact on the features of such devices can be considered substantial and represents an important issue of nanoscale system architecture. Therefore fabrication and planning techniques could benefit from a theoretical understanding of the electronic transport properties of CNTs, not in their infinite or electrically isolated form, but in the metal-nanotube-metal structure.

CNTs have been classified initially by early electronic band structure calculations [4-7] and afterwards by scanning tunnelling microscopy [8-9] as metals, semimetals or semiconductors on the strict basis of their diameter and helicity [10]. Not all valence electrons contribute to this particular conducting behaviour, since the three $\sigma$-electrons of the valence states that form bonds along the cylinder walls have bands that are too far away from the Fermi level [11]. On the other hand, the bonding and antibonding $\pi$ and $\pi^*$ bands perpendicular to the CNT's surface characterise electronic properties near the Fermi level. Therefore it is a common practise to ignore the contribution of $\sigma$-electrons in the calculation of CNT electronic transport (Tight-Binding approach [12]). However, this simplification seems insufficient in the description of bonds formed between the metallic contact and the carbon atoms, or in the case of $\sigma$-$\pi$ mixing provoked by wall curvature effects. First principle calculations [13] solve such complications describing in a realistic way all bonding interactions but have the handicap of being computationally expensive. On the basis of these assumptions there have been various approaches up to now regarding transport calculation of finite carbon nanotube systems. Some address finite geometries simulating contact effects with an addition of a simple appropriately parameterised self-energy term in Green's functional calculation [14]. Such approaches neglect a quantitative description of the chemical bond between carbon and metal atoms. Others address the problem with a realistic contact description using *ab initio* [13] or extended Hückel calculations [15], for a limited number of CNTs and contact types. However, a systematic study of interfacial reconstruction effects on the conduction mechanism, varying the CNT helicity and the contact-CNT interface geometry is still lacking.

In our approach we use the Non Equilibrium Green's Function formalism [16] with two diverse Hamiltonians for the transport calculation of gold-CNT-gold systems. The first one, i.e. the Tight Binding (TB) Hamiltonian, considers only $\pi$-electrons, while the second, i.e. the Extended Hückel (EH) one, includes all valence electrons. The two models differ in the approach of the chemical bond. Indeed, the TB model is a typical model Hamiltonian where the bond strength does not depend on the curvature and the geometric arrangement. Therefore, in the case of CNTs, the local bonding topology in the TB model is equivalent to the bonding topology of the 'flat' graphene sheet. As a consequence, the energy bands of infinite CNTs calculated with the TB Hamiltonian are in agreement with the classification derived by folding the graphene two-dimensional energy band (zone folding approximation) [11]. In turn, the EH model considers the effects of the actual geometric configuration on the bond strength allowing, in the meanwhile, extensive calculations in relatively large systems, since it is considerably less computationally expensive with respect to the *ab initio* approaches. Moreover, we note that EH, nonetheless a semiempirical method, has been proven to give results close to other first-principle calculations in the case of a (6, 6) armchair nanotube [17].

The aim of this work is twofold. Firstly, we calculate the transmission function of various gold-CNT-gold systems with different nanotube helicities using both TB and EH Hamiltonians, investigating the interference between the CNT conduction channels caused by the contacts. Moreover, comparing the TB and EH results, we identify the application limits of the TB approach, which in turn allows very large-scale quantum transport calculations. Secondly, using thereon the EH method, we focus on the role of the relative geometric arrangement between the gold lead and the CNT as well as that of electron charging on the conduction mechanism for the cases of both metallic and semiconducting CNTs.

The paper is structured as follows: In section 2 we give a theoretical description of Green's formalism with the respective TB and EH Hamiltonians. In section 3 the TB and EH methods are confronted and various transmission functions of different nanotube systems are demonstrated. Section 4 focuses on the effect that small variations in the tube-contact interface geometry provoke on the conduction mechanism. Section 5 analyses the effect of electron charging on transport. Finally in section 6 we discuss our results.

## 2. Computational methodology

We use the Non-Equilibrium Green Functional formalism in order to study the quantum transport of finite size CNTs "sandwiched" between two semi-infinite gold contacts (i.e. the two-terminal geometry). Our approach is based on the single particle retarded Green's function matrix

$$G = \left[ ES - H - \Sigma_L - \Sigma_R \right]^{-1} \tag{1}$$

where $H$ is the 'device' Hamiltonian in an appropriate basis set, $S$ is the overlap matrix in that basis set, $\Sigma_{L,R}$ is the self energy which includes the effect of scattering due to the left ($L$) and right ($R$) contacts. The contact self-energy can be expressed as $\Sigma = \tau g_s \tau^\dagger$ where $g_s$ is the Green's function of the contact restricted to the surface zone and $\tau$ is the hamiltonian relative to the mutual interaction between the CNT and the contact. In the case of coherent transport, the current can be calculated directly using the Landauer-type expression

$$I = \frac{2e}{h} \int_{-\infty}^{+\infty} dE \, T(E) \left[ f(E, \mu_L) - f(E, \mu_R) \right] \tag{2}$$

where the transmission is

$$T(E) = Tr \left[ \Gamma_L G \Gamma_R G^\dagger \right] \tag{3}$$

with

$$\Gamma_{L,R} = i \left[ \Sigma_{L,R} - \Sigma_{L,R}^\dagger \right].$$

In equation 2 $f(E, \mu_{L,R})$ is the Fermi-Dirac distribution of electrons in the contact at chemical potential $\mu_{L,R} = E_F \pm 0.5V$ [16,18] where $E_F$ is the Fermi energy of the system and $V$ the applied potential.

Charging effects must be considered when an external bias $V$ is applied to the CNT (i.e. in non-equilibrium conditions) since the total number of electrons in the device varies due to the flow of charge from the contacts. Charging effects can be computed with the inclusion of a self consistent potential $U_{SC}\{\rho\}$ in the formalism, which is a functional of the density matrix $\rho$ in the device and depends also on the external bias. The device Hamiltonian now reads

$$H = H_0 + U_{SC}\{\rho\}$$

where $U_{SC}\{\rho\}$ must be determined self consistently and $\rho$ is given by the expression below.

$$\rho = \frac{1}{2\pi} \int_{-\infty}^{+\infty} dE \left[ f(E, \mu_R) G \Gamma_R G^\dagger + f(E, \mu_L) G \Gamma_L G^\dagger \right] \qquad (4)$$

Appropriate Pariser-Parr-Pople (PPP) type functional forms of $U_{SC}\{\rho\}$, depending on a set of semi-empirical parameters, have been introduced in the TB and EH models [18,19]. However, the potential profile inside a sufficiently long CNT could be assumed to be flat with the voltage drop largely occurring at the contacts. This assumption is confirmed experimentally by potentiometric measurements of potential profile [20], and by early calculations based on PPP type potentials [21]. That being the case, we can approximate $U_{SC}\{\rho\}$ as [16]

$$U_{SC}\{\rho\} \approx U_0(N - N_{eq}) \qquad (5)$$

where $U_0$ is the mean charging energy per electron, $N$ ($N_{eq}$) the total number of electrons present in the device in non-equilibrium (equilibrium) conditions, which can be calculated using (4). In this case charging is described by the single empirical parameter $U_0$. Note that the constant potential (5) merely causes a shift of the device energy levels, thus it affects all the related energy dependent quantities, like $T(E)$ or DOS, only for a re-scaling of the energy.

In order to describe the CNT and contacts electronic behaviour, we use two semiempirical Hamiltonian matrixes, which are formally distinguished by the choice of basis set function: the Extended Hückel Model and the Tight Binding model.

The EH model uses all valence orbitals which are approximated with Slater type orbital (non-orthogonal) functions. The single particle Hamiltonian matrix elements are

$$H_{i,j} = \begin{cases} -V_i & if \ i = j \\ \dfrac{S_{ij}}{2}\left(V_i + V_j\right) & if \ i \neq j \end{cases}$$

here the indices $i$ and $j$ run over the valence orbitals. The overlap matrix $S_{ij}$ and the diagonal elements $V_i$ can be calculated, given the system geometry, using the appropriate parameterisation [16,22].

In the Tight Binding model the basis functions are the (orthogonal) $p_z$ orbitals centred at each atom and the single particle Hamiltonian matrix elements are

$$H_{i,j} = \begin{cases} \varepsilon_0 & if \ i = j \\ -t & if \ i,j \ are \ next \ neighbors \\ 0 & otherwise \end{cases}$$

Here the indices $i, j$ run over the atom and the constants $\varepsilon_0$ and t are 0 and 2.66eV, respectively, for the inner carbon atom of the CNT [23]. For a reliable comparison of the results obtained by the two models, the hopping integral $t_{Au\text{-}Au}$ between two gold atoms has been set as so in order to produce the same surface Density of States at the Fermi level by both TB and EH models. The hopping integral $t_{C\text{-}Au}$ between a gold and a carbon atom has been considered as a free parameter and its determination will be discussed below. We note that, while the EH method allows a direct evaluation of the geometrical configuration between the CNT and the pad and internally determines the bonding scheme, in TB the contact bonding reconstruction is quite arbitrary. Indeed, here we need to use a distance cutoff $d_{cutoff}$ rule in order to determine if bonds are formed between C and Au atoms at the extremes of the CNT. Moreover, we impose that bonds in the TB model are characterised by the same value of the hopping integral $t_{C\text{-}Au}$, even if the distances between the C and Au atoms forming the bonds are different.

For our transport calculations we considered defect-free single-walled carbon nanotubes of the armchair and zigzag types with different lengths and diameters. In the case of zigzag CNTs we considered both metallic and semiconducting structures. At the dangling bonds of the two ends of these finite CNTs we attached the Au(111) contacts (figure 1a). Depending on the objectives that were studied, we performed our simulations for different contact-nanotube geometrical configurations that varied the distance and the symmetry of the contacts with respect to the tube.

## 3. Finite size metal-CNT-metal conductance: TB and EH results

In order to compare the transport properties estimated using the EH and TB methods we performed a series of calculations for identical system configurations using both models. Metallic and semiconducting CNTs of diverse dimensions and helicities were examined. Nanotube lengths were chosen as so in order to minimise the influence of introduced energy states by the metallic leads around $E_F$ [24], avoiding excessive lengths that could provoke a computational load. Where possible, pads were placed symmetrically with respect to the tube's endings, usually having each C atom at the two extremes of the CNT ideally forming (generally not equivalent) bonds with 3 Au atoms in a 'stick and ball' geometrical representation. In figure 1b the contact for the (12,0) case is shown, where a highly symmetric CNT-pad configuration can be obtained. The other contact atomic configurations considered in this section are shown in figure 2. We note that, whereas the C atoms in the last ring are not perfectly positioned on a hollow position, most of them have 3 neighbouring Au atoms. The (5,5) case is a relevant exception since C atoms in the second ring of the nanotube are nearest to some Au contact atoms with respect to the atoms of the first nanotube ring. As we will see in the following, a result of this peculiar geometric arrangement is a strong difference of the transport properties calculated by the two models. In the calculations of the current-voltage characteristics shown in this and in the next section charging effects are neglected and the $E_F$ has been set to the value of the respective infinite CNT. The influence of charging effects and possible $E_F$ variations on the conductance of our metal-CNT-metal systems is discussed in section 5.

Figures 3a and 3b show as solid lines the transmission spectra obtained using the two models for the case of a (9,0) finite carbon nanotube formed by 8 unit cells (~ 33.3 Å long). Transmission lines of infinite systems are also plotted in the graphs as dashed lines. The contact-nanotube distance was established at 1.0 Å [15] and maintained as so for all results presented under this section. In addition, all graphs of this manuscript were rescaled in order to refer to the Fermi energy $E_F$ of the respective infinite system as zero energy. In the calculation based on the TB Hamiltonian the distance cutoff was set as $d_{cutoff} = 2.5$ Å. In the case of the TB graph in particular, two spectra are shown obtained using two values of the $t_{C-Au}$ hopping integral: $t_{C-Au} = 0.45 t_{Au-Au}$ (solid line) and $t_{C-Au} = 0.3\ t_{Au-Au}$ (points). The corresponding current-voltage (I-V) characteristics calculated from the three transmissions plotted in figure 3 are shown in figure 4.

The step like behaviour of the transmission (conductance quantization) is expected for the infinite (ideally metal contacted) CNT due to its quasi one-dimensional character. Moreover, in the case of a metal nanotube the zone-folding approximation (which is equivalent to the TB model in the case of an infinite nanotube) predicts a conductance value of $2 \times G_0$ ($G_0 = 2e^2/h$) near the Fermi Energy due to the presence of two spin degenerate bands at the Fermi level [11]. The same $2 \times G_0$ conductance value is likewise predicted by the EH method for an infinite nanotube near $E_F$, apart from a small secondary gap present at $E_F$. This secondary gap is a characteristic of metallic zig-zag nanotubes and can be evidenced only when the band structure is calculated by means of theoretical approaches beyond the zone-folding approximation (e.g. using *ab initio* Hamiltonians) [11]. The overall transmission behaviour of a finite size nanotube is irregular due to both the finite-size effect (where a discrete set of eigenvalues replaces the continuous bands) and the effects of coupling with the contact (i.e. broadening of peaks and lifting of degeneracy). In particular, the plateau at $2 \times G_0$ near the Fermi level is replaced by peaks at $2 \times G_0$, which are present for both models, while peaks in energy regions away from the Fermi level remain well below the conductivity plateaux of an infinite nanotube. The average conductance value calculated by the EH model in a 2V range across the V=0 value is $1.38 \times G_0$. However, for finite nanotubes the structure of the transmission spectra is model dependent even near the Fermi level, indicating a strong dependence of the transport properties on the chemical bond reconstruction of the CNT-metal interface. Moreover, in the case of the TB model the $t_{C-Au}$ value affects the peaks broadening, although maxima are anyway at $2 \times G_0$.

These results demonstrate that the simplified (on-off) approach to the interfacial bond problem in the TB model is, of course, inadequate for an accurate study of the transport in a metal-finite CNT-metal system. However, we can try to get an optimal estimate of the $t_{C-Au}$ in order to obtain a similar estimate of the I-V characteristic trend for both models. We found that using the value $t_{C-Au}=0.45 \times t_{Au-Au}$ we obtain a good agreement between the I-V curves calculated by means of the two models (see figure 4) for a large range of the potential across the V=0 value. In order to determine this optimal value of the $t_{C-Au}$ for a contact distance of 1 Å we have considered the CNTs that present a metallic character for both TB and EH models. Of course, any attempts of recovering similar transport features by the two methods fail when the divergence between TB-and EH-based results is noteworthy (e.g, see the following, when one model predicts a semiconductor character while the other a metallic one, or when a channel block appears only for one of the two models). In figure 4 we can also appreciate the deviation of the I-V when a $t_{C-Au}$ value of $0.3 \times t_{Au-Au}$ is used, which differs from the optimal value for about 33%. The $t_{C-Au}=0.45 \times t_{Au-Au}$ value is used in this section for all calculations based on the TB model.

The discrepancy between TB and EH is more fundamental in the case of nanotubes with smaller diameters. Indeed, it is well known that zone folding approximation is no longer valid for nanotubes with radii lower than ~ 3.5 Å [12] and the usual classification in terms of metallic and semiconductor CNTs cited above looses any meaning. The *de-facto* more accurate and sophisticated EH method gives results that differ, since it incorporates σ-π rehybridisation effects due to wall curvature on the calculation of the system transmission function. In the set of figure 5 we show as solid lines the transmission spectra calculated using the EH model for the case of (2,2), (3,3), (4,0) and (5,0) finite CNTs. The average conductance values calculated in the 2V range across V=0 are $0.96 \times G_0$, $0.4 \times G_0$, $0.43 \times G_0$, and $0.62 \times G_0$ respectively. Transmissions of the infinite nanotubes are also shown as dashed lines. An analogous analysis of the transmission of small radius nanotubes obtained using the TB model is shown in figure 6. The transmission calculation for an infinite nanotube based on the TB model gives results in agreement with the usual classification and therefore (2,2) and (3,3) nanotubes are metallic with a $2 \times G_0$ plateau while (4,0) and (5,0) nanotubes have a semiconducting character. EH estimates show that the TB approximation is inadequate for the transport calculation of small radius nanotubes. Indeed, the transmission plots shown in figure 5 demonstrate that all nanotubes considered have a metallic character. The transmission value in the $E_F$ region is $2 \times G_0$ only for the (3,3) nanotube while it is $G_0$ or $3 \times G_0$ for all the others. A similar transmission behaviour can be deduced by *ab initio* band calculations performed in the same structures [25-26], confirming the good approximation of EH results with the respective first-principle ones. In the case of (2,2), (4,0) and (5,0) nanotubes the transmission calculation for a finite system shows the typical irregular structures of a finite system with a sequence of peaks which reaches the $G_0$ plateaux of the infinite system in the $E_F$ region, but never reaches the $3 \times G_0$ plateaux. This feature indicates that contacts provoke interference between the states at energy levels relative to three bands, while when a single band is present no blocking effect occurs.

In turn, in the case of the (3,3) nanotube the transmission peaks of the finite system reach $G_0$ instead of $2 \times G_0$ and this characteristic is observed for both Hamiltonian models used in the calculations (see figures 5, 6). This fact indicates the blocking of one of the two conductance channels and is related to a lack of coupling with the electrodes for orbitals with a given symmetry (see Ref. [27] for a discussion on this issue based on the TB model). This diversion from the $2 \times G_0$ can be justified as a contact and not as a CNT effect, since the agreement of the two models implies some interrupted transport channel due to a particular pad-tube geometry. Moreover, as evidenced by both methods, blocking is inherently related to the geometrical symmetry of the nanotube-contact interface and it is not merely a scattering effect.

Infinite conduction analysis of large radius nanotubes (see figures 7,8) shows that TB and EH predictions qualitatively agree in the zone near the Fermi energy, confirming the reliability of the zone folding approximation for nanotubes with radii greater that ~3.5 Å. In fact, the (9,0) zigzag system demonstrates a ~ $2 \times G_0$ conductance, while (10,0) and (11,0) semiconductors confirm energy gaps, with the smaller in diameter nanotube giving a wider gap, as expected. However, the breadth of the energy gap calculated by the TB model, does not always coincide with results obtained by the EH method. A blocked conduction channel is observed in the case of the (6,6) armchair CNT with a similar behaviour of that discussed in the case of the (3,3) CNT. The (5,5) armchair CNT presents a singularity since an interrupted transport channel is evident only in the case of the TB model. However, the particular geometry of this system implies that atoms in the second layer of the nanotube are nearest to some Au contact atoms with respect to atoms in the first nanotube layer. This fact has a noteworthy influence in the case of TB where a cut-off rule is applied in order to determine the bond occurrence. On the other hand, EH does not present any contact interference for this tube and transmission peaks arrive at $2 \times G_0$. The importance of the channel blocking effect by the metallic contact can be also appreciated in the calculation of the mean conductances for the CNTs mentioned previously in a 2 V zone around V=0, where for the (5,5) tube we get a higher value ($1.25 \times G_0$) than for the (6,6) one ($1.06 \times G_0$). Finally, in all cases -also those where the TB and EH methods appear to agree- there always exist small-scale disparities that can be important for currents of the order of μA.

## 4. Contact geometry dependence of transport properties

Simulations were also performed in order to study the influence of electrode positioning with respect to the nanotube on the electronic transport of finite CNT systems. For these calculations we preferred the EH Hamiltonian to its strictly π-electron TB counterpart, taking into account the considerations discussed in the previous paragraph, in particular the descriptive capacity of the EH model when it comes to Au-C chemical bonding. To begin with, we considered two 8-unit-cell CNTs (a (9,0) and a (10,0)), which we 'sandwiched' between the two gold pads, using at first the same symmetrical configuration between C and Au interface atoms as above. We initially focused on the contact-nanotube distance effect on transport, calculating the transmission spectra for four diverse snapshots of our systems with a successive variation of this distance from 2 to 0.5 Å. We have to note here that an optimal (minimum energy) distance between the pad and the nanotube can be estimated theoretically using a Hamiltonian that includes the nuclear contribution other than the electronic one (applying e.g. *ab initio* models). However, different models or different approaches to the boundary conditions (i.e. periodic, constant pressure etc) at the termination of the computational box end up in calculating different optimal distances, which deviate in a significant manner. Moreover, it is unlike that an eventual experimental implementation of the metal-nanotube-metal device leads to a real contact-nanotube distance equal to the optimal one. Evaluating the latter, we have concluded that a more appropriate and more consistent (in the framework of pure electron Hamiltonians) study of the distance effect can be confronted by calculating the transport properties varying the interface geometry for a wide range of contact-nanotube distances, including also the optimal one.

Figures 9 and 10 show simulation outcomes obtained. The resulted transmission curve for the (9,0) system reveals poor conductance for a 2-Å contact-tube distance (shorter peaks around $E_F$), while when bringing the pad closer a respective increase in the current flow is observed with peaks arriving at $2 \times G_0$. The mean conductances (see figure 11) calculated for a 2 V range across V=0 are $0.48 \times G_0$ (2 Å), $1.13 \times G_0$ (1.5 Å), $1.38 \times G_0$ (1 Å) and $1.53 \times G_0$ (0.5 Å) respectively, supporting previous observations. A more particular result is obtained in the case of the (10,0) system where a conductance gap is expected in the Fermi zone. Here, a 2-Å contact-tube distance contrarily reveals small transmission peaks around $E_F$. This singularity can solely be described as a weak-coupling effect between interface C and Au atoms. In fact, these peaks are not present either in the case of the infinite (10,0) nanotube (figure 7), or when the contact-CNT distance is smaller. Therefore, they can only be attributed to those C atoms at the extremes of the CNT that are not capable of forming regular bonds with atoms of the gold contact and thus introduce undesirable energy states near the Fermi level. As we can see in figure 10, when bringing the contacts closer, peaks slowly disappear and the energy gap is being formed. Analogous considerations can be sustained from the analysis of the I-V curves of the mentioned systems (figures 11, 12). Acquired results demonstrate an important influence of the distance factor between nanotubes and metallic electrodes, conditioned by the strength of interface couplings. Fabricators of nanotube devices with Au contacts that follow similar geometries have to take into consideration that for distances as big as 1.5 Å (when weak couplings start to be observed) unexpected behaviours may emerge.

Up to this point all simulations presented in this manuscript considered an ideal symmetrical interface representation schema between C and Au atoms. For a topological approximation more likely probable in laboratory conditions we also examined the conducting behaviour of finite CNTs under no symmetrical contact-CNT geometrical configurations. For these measurements we considered the systems described above, with a fixed 1-Å pad-tube distance, but with the symmetry between them altered. Various "twists" of one or both contacts (see figure 13b) as well as an arbitrary rotation and displacement of the metallic pad with respect to the centre of the CNT (see figure 13c) have been considered. Obtained transmission functions (figure 14) reveal small modifications in the spectra. Some peaks appear with altered heights and can be attributed to slightly modified interference effects provoked by the new geometry, similar to those discussed in the previous section. Equivalently, moderately shifted peak positions and broadenings are contact generated effects. The maximal variation from the 'symmetric' positioning can be observed for the last contact configuration considered. However, the overall behaviour of the systems remains unaltered, leading to the assumption that a contact symmetry effect is present but plays a marginal role in the electronic transport of metal-CNT-metal systems.

Finally, in order to exclude that previous considerations are intrinsic only to the two types of nanotubes studied above, we expanded our investigation to various gold-CNT-gold systems (including the systems studied in section 3), repeating same type of calculations. Results confirmed that the behaviour met in the two previous systems can be characterised as typical, since the distance factor is significantly correlated with the conducting capacity of the systems, whereas the symmetry one has always an impact of secondary importance.

## 5. Electronic charging influence on transport

Electron charging effects can be evidenced in metal-nanotube-metal systems a) under no bias due to the difference between the electrochemical potentials of the device and the electrodes, or b) under bias due to the electron flow from and out of the contacts. In equilibrium conditions (no bias), charging is inherently related to the relative position of the system's (pads+nanotube) Fermi energy with respect to the CNT's energy levels and consequently with the formation of Schottky type barriers. The exact location of $E_F$ is controversial even if *ab initio* approaches are applied [28]. In the spirit of the semi-empirical approaches applied in this work it is more appropriate to consider the $E_F$ level as a free parameter, which should vary approximately between the Fermi energy of the infinite CNT and that of the contacts [16]. An empirical estimate of the $E_F$ level can be obtained eventually by fitting experimental results [18]. Anyhow the corrections in the I-V characteristics due to charging effects under bias conditions can be evaluated by means of our approach, while in consistence to the previous discussion we have considered different values of the $E_F$ level in our calculations. We have applied the approximations discussed in section 2 and fixed the CNT's one electron charging energy $U_0$ to 2,94 eV (this value can be derived in the framework of the complete neglect of differential orbital theory using experimental data for the carbon ionization potential and the electron affinity [29]). The EH hamiltonian has been applied for these calculations.

Figure 15 shows I-V curves obtained for a (9,0) and a (10,0) CNT before (blue line) and after (red line) the correction induced by the self-consistent potential for three values of the $E_F$. The distance between the pad and the nanotube is 1 Å and a symmetric positioning is considered. The dependence of the I-V characteristics on $E_F$ can be understood looking at the transmission spectra of the two systems shown in figures 3 and 7 respectively. In particular, we should expect a strong variation of the I-V curve for the (10,0) case when the $E_F$ value is outside the gap, i.e when $E_F$=-9,5 (the value of the gold contact). It is interesting to note that the I-V correction for both CNTs becomes significant only if we consider the system's electrochemical potential close to the Fermi energy value of the metal contact, whereas when this lies near the infinite CNT's Fermi energy, no important change can be observed. This feature is related to the less symmetric position of $E_F$ with respect to the transmission spectra (and consequently to the density of states spectra) when $E_F$ moves away from the value of the infinite CNT (see again figs. 3 and 7). Indeed, the effect of charging is more relevant if one contact is preferentially correlated to the device states with respect to the other contact. The applied potential makes the chemical potential of the contact cross theses states and, consequently, charges move from the contact to the device without a comparable charge transfer form the device to the other contact. The self-consistent potential reacts to the device charging and its effect is a displacement with respect to the contact chemical potential of the energy levels related to these states. Thus, the self-consistent potential effect causes a less pronounced increase of I with respect to V until the chemical potential related to the other contact reaches the states in the opposite region of the spectra (this condition occurs at about ± 2-2.5 V for the $E_F$=-9.5 eV case). Contrarily, when $E_F$ is positioned symmetrically with the states present for $E<E_F$ and $E>E_F$ a balanced charge transfer between the device and the two pads occurs reducing sensibly the charging effects [16,18].

## 6. Discussion

In this manuscript we have reported our investigations on the electronic transport properties of finite carbon nanotubes attached to gold metallic contacts. Calculations were performed using the non-equilibrium Green functional formalism with a Tight-binding and an Extended Hückel Hamiltonian. We have focused our study on the effects of the interfacial arrangement on the conduction properties of the metal-nanotube-metal systems. Charging and its relationship with the problem of Fermi energy positioning has also been discussed. Contact interference effects have been observed for small radius nanotubes between the states at energy levels relative to three bands, while when a single band is present no blocking effect occurs. Blocked conduction channels at two-band energy levels have been noted for (3,3) and (6,6) CNTs and likewise for (6,0) and (12,0) CNTs (not shown), whereas no such effect has been calculated for (5,5) and (9,0) CNTs. From our extensive analysis we can conclude that the chemical nature of the electrode is not the only factor which rules contact transparency. Indeed, by fixing the electrode type and changing CNT helicities we have observed a complex variety of effects due to band mixing and interference induced by the contact. As a practical consequence we point out that there should not be a single metal option for making high transparency contacts to CNTs, but the best choice ever depends on the helicity of the CNT to be contacted.

It is likely that the experimental preparation of metal-CNT contacts does not lead to any optimal interface reconstruction or to any symmetrical arrangement between the relative positions of the C and metal atoms. In this sense exploring the conduction properties varying the contact geometry is more correct. Therefore, instead of focusing on an optimised structure, we have examined the effect of the interfacial geometrical configuration between the contact and the CNT on the conduction mechanism and found that contact-nanotube distance has a significant impact, while the role of the contact-nanotube symmetry is marginal, but still evident. We can speculate that fabrication of metal-CNT-metal devices should be affected mostly by a symmetry variance; therefore we should expect quite homogeneous conductance properties among equivalent (same metal, same size, same helicity) devices. However a strong distance variance should arise when a contact is represented by a metallised atomic force microscope tip used as scanned electrical nanoprobe in conductance measurements [30]. In this case our results indicate that a careful analysis should be performed in order to understand the experimental results.

We have furthermore confronted results obtained by both EH and TB methods and found that the TB approximation is inadequate for the transport calculations of small radius nanotubes or when a realistic description of the contact effect has to be considered. Our study, therefore, demonstrates the necessity to consider a realistic description of the metal-device interface connectivity in the computer-assisted design of molecular devices. More generally the use of model Hamiltonians like TB, although being useful in order to recover some significant features of the quantum transport, are inappropriate when the chemical bond plays a significant role in controlling the conduction properties of the molecular system. Anyhow, a sequential multi-scale approach can be still applied when the model Hamiltonian reproduces the most significant transport aspects of the nano-structure, as in the case of metal contacted large radius CNTs. In this case free parameters in the model Hamiltonian can be optimally estimated using more accurate schemes in small-size systems, allowing reliable transport calculations in large-size systems.

Finally we note that, although we concentrate our attention on the contact effects over the transport features (which should be important when small size CNTs are considered), we are aware that other scattering (electron-phonon, electron-electron) mechanisms can strongly influence the conductance of these systems [14]. Future extensions of this work will be devoted to address the problem of a realistic and reliable modelling of both scattering and contacts in CNT based systems.

The authors would like to thank P. Alippi for the many useful discussions and V. Privitera, S. Scalese who indicated the experimental relevance of the problem in study.

**LIST OF FIGURE CAPTIONS**

**Figure 1.** a) Geometrical representation of the studied gold-nanotube-gold system. Contacts are considered as semi-infinite. b) 'Stick and ball' symmetry between interface C and Au atoms of a (12,0) nanotube and a (111) gold contact, where each C atom is positioned within three Au atoms.

**Figure 2.** Contact-CNT interface configurations of the systems used for the transport calculations of section 3.

**Figure 3.** Transmission as a function of energy for a (9,0) 8-unit-cell CNT calculated with a) an EH and b) a TB Hamiltonian. Transmission lines of infinite systems are also shown in the graphs as dashed lines. The Fermi energy of the respective infinite CNT is set to zero, while the black point on the Energy axis of the EH transmission spectra indicates gold Fermi energy ($E_{F_{Au}} = -9.5eV$). In the case of TB two spectra are shown using two values of the $t_{C-Au}$ hopping integral: $t_{C-Au} = 0.45 t_{Au-Au}$ (solid line) and $t_{C-Au} = 0.3\ t_{Au-Au}$ (points). The average conductance value calculated in the 2V range across V=0 in the case of EH is $1.38 \times G_0$, whereas in that of TB it is $1.31 \times G_0$, when $t_{C-Au} = 0.45 t_{Au-Au}$, and $0.74 \times G_0$, when $t_{C-Au} = 0.3\ t_{Au-Au}$

**Figure 4.** I-V curve for a (9,0) 8-unit-cell CNT with a contact distance of 1 Å, calculated with the EH method (solid line) and the TB method using two hopping integrals: $t_{C-Au}$=0.45 $t_{Au-Au}$ (blue line) and $t_{C-Au}$=0.3 $t_{Au-Au}$ (red line).

**Figure 5.** Transmission as a function of energy for a 20-unit-cell (2,2), a 15-unit-cell (3,3), a 10-unit-cell (4,0) and an 8-unit-cell (5,0) CNT, calculated with the EH model. Transmission lines of infinite systems

are also shown as dashed lines. Fermi energy of the respective infinite CNTs is set to zero. Black points on the Energy axis denote Au Fermi energy. The average conductance values calculated in the 2V range across V=0 are 0.96×$G_0$, 0.4×$G_0$, 0.43×$G_0$, and 0.62×$G_0$ respectively.

**Figure 6.** Transmission as a function of energy for a 20-unit-cell (2,2), a 15-unit-cell (3,3), a 10-unit-cell (4,0) and an 8-unit-cell (5,0) CNT, calculated with the TB model. Transmission lines of infinite systems are also shown as dashed lines. The average conductance values calculated in the 2V range across V=0 are 0.99×$G_0$, 0.39×$G_0$, 7×$10^{-3}$×$G_0$, and 8×$10^{-6}$×$G_0$ respectively.

**Figure 7.** Transmission as a function of energy for a 16-unit-cell (5,5), a 14-unit-cell (6,6), an 8-unit-cell (10,0) and an 8-unit-cell (11,0) CNT, calculated with the EH model. Transmission lines of infinite systems are also shown as dashed lines. Fermi energy of the respective infinite CNTs is set to zero. Black points on the Energy axis denote Au Fermi energy. The average conductance values calculated in the 2V range across V=0 are 1.25×$G_0$, 1.06×$G_0$, 0.07×$G_0$ and 0.17×$G_0$ respectively.

**Figure 8.** Transmission as a function of energy for a 16-unit-cell (5,5), a 14-unit-cell (6,6), an 8-unit-cell (10,0) and an 8-unit-cell (11,0) CNT, calculated with the TB model. Transmission lines of infinite systems are also shown as dashed lines. The average conductance values calculated in the 2V range across V=0 are 0.89×$G_0$, 0.73×$G_0$, 6.5×$10^{-2}$×$G_0$, and 1.5×$10^{-2}$×$G_0$ respectively.

**Figure 9.** Transmission as a function of energy for four snapshots of a (9,0) 8-unit-cell system with a varying contact distance of 2 Å, 1.5 Å, 1 Å and 0.5 Å. The respective mean conductances for a 2 V range across V=0 are: 0.48×$G_0$, 1.13×$G_0$, 1.38×$G_0$ and 1.53×$G_0$.

**Figure 10.** Transmission as a function of energy for four snapshots of a (10,0) 8-unit-cell system with a varying contact distance of 2 Å, 1.5 Å, 1 Å and 0.5 Å. The respective mean conductances for a 2 V range across V=0 are: 0.11×$G_0$, 0.1×$G_0$, 0.07×$G_0$ and 0.06×$G_0$.

**Figure 11.** I-V functions for a (9,0) 8-unit-cell system with a varying contact distance of 2 Å, 1.5 Å, 1 Å and 0.5 Å. The respective mean conductances for a 2 V range across V=0 are: 0.48×$G_0$, 1.13×$G_0$, 1.38×$G_0$ and 1.53×$G_0$.

**Figure 12.** I-V functions for a (10,0) 8-unit-cell system with a varying contact distance of 2 Å, 1.5 Å, 1 Å and 0.5 Å respectively. The respective mean conductances for a 2 V range across V=0 are: 0.11×$G_0$, 0.1×$G_0$, 0.07×$G_0$ and 0.06×$G_0$.

**Figure 13.** Contact-nanotube interface configurations used to study the effect of symmetry on transport for a (9,0) (upper line) and a (10,0) (lower line) CNT. a) 'Standard' configuration used throughout this article. b) Anti-clockwise rotation of the contact with respect to the CNT by 10° and 20° for the (9,0) CNT, and by 9° and 18° for the (10,0) CNT. c) An 'arbitrary' rotation and movement of the contacts with respect to the centre of the CNTs (the two figures per CNT represent left and right contacts).

**Figure 14.** Transmission as a function of energy for a (9,0) and a (10,0) CNT with altered contact-tube geometrical configurations with respect to previous calculations (grey fillings). a) (9,0) CNT with one contact rotated in an anti-clockwise manner by 10° (blue line) and 20° (red line). The respective mean conductances for a 2 V range across V=0 are $1.36 \times G_0$ and $1.42 \times G_0$. b) (10,0) CNT with one contact rotated in an anti-clockwise manner by 9° (blue line) and 18° (red line). The respective mean conductances for a 2 V range across V=0 are $6.2 \times 10^{-2} \times G_0$ and $6.8 \times 10^{-2} \times G_0$. c) (9,0) CNT with both contacts rotated by 10°, -10° (blue line) and 10°, -20° (red line) respectively. The respective mean conductances for a 2 V range across V=0 are $1.28 \times G_0$ and $1.39 \times G_0$. d) (10,0) CNT with both contacts rotated by 9°, -9° (blue line) and 9°, -18° (red line) respectively. The respective mean conductances for a 2 V range across V=0 are $5.9 \times 10^{-2} \times G_0$ and $6.7 \times 10^{-2} \times G_0$. e) (9,0) CNT with both contacts rotated and moved according to figure 13c (upper). The respective mean conductance for a 2 V range across V=0 is $1.39 \times G_0$. f) (10,0) CNT with both contacts rotated and moved according to figure 13c (lower). The respective mean conductance for a 2 V range across V=0 is $6.7 \times 10^{-2} \times G_0$.

**Figure 15.** I-V functions for a (9,0) and a (10,0) CNT before (blue line) and after (red line) the correction imposed by the application of the self-consistent potential. The Fermi energy value of the infinite (9,0) and (10,0) CNTs are $E_F$=-10.29 eV and $E_F$=-10.33 eV respectively, whereas that of the gold contact is $E_F$=-9.5 eV.

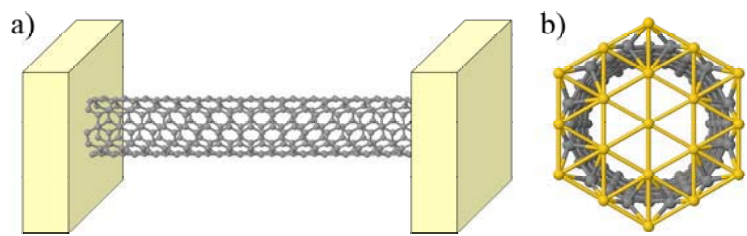

**Figure 1**

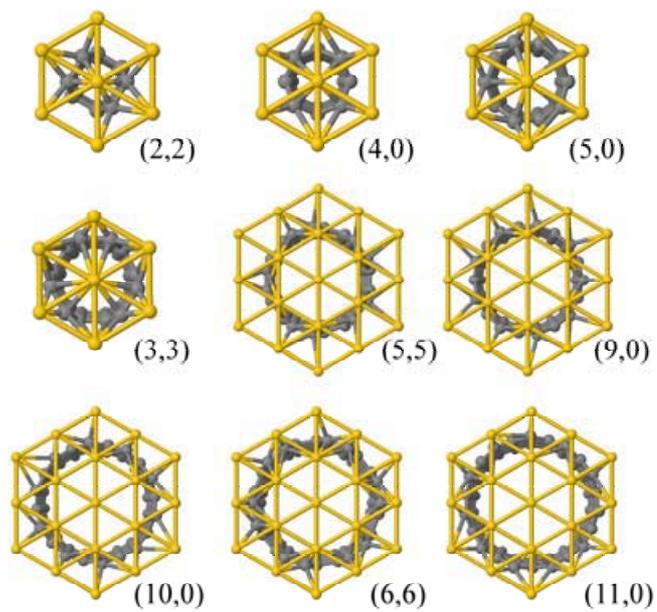

**Figure 2**

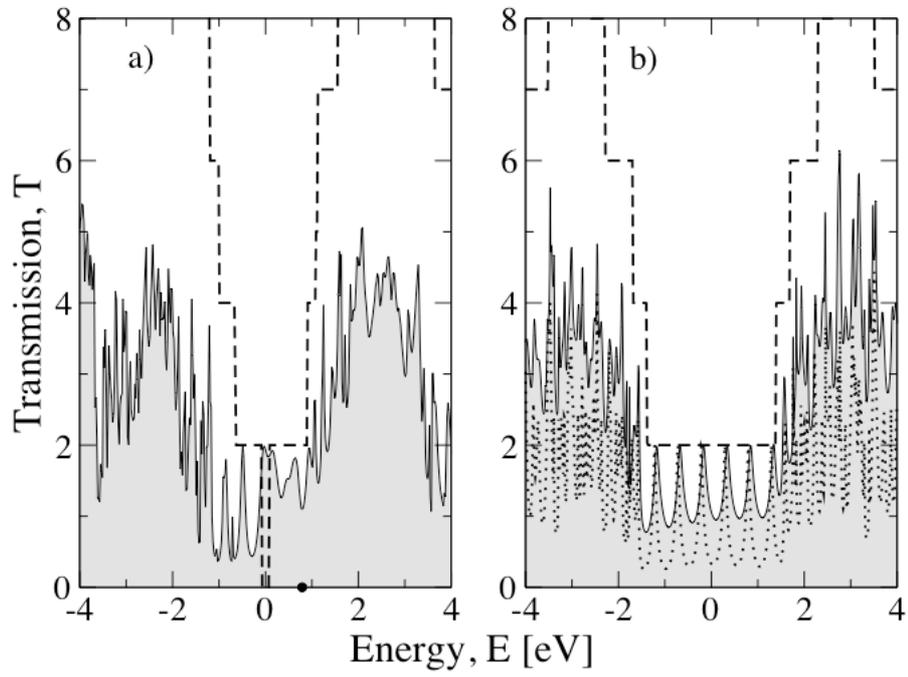

**Figure 3**

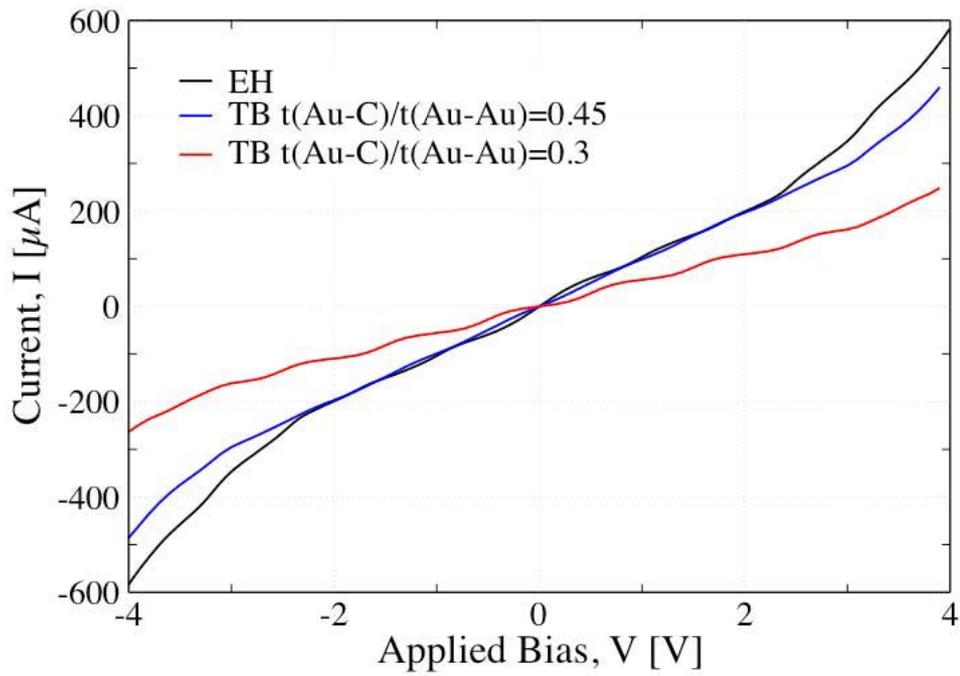

**Figure 4**

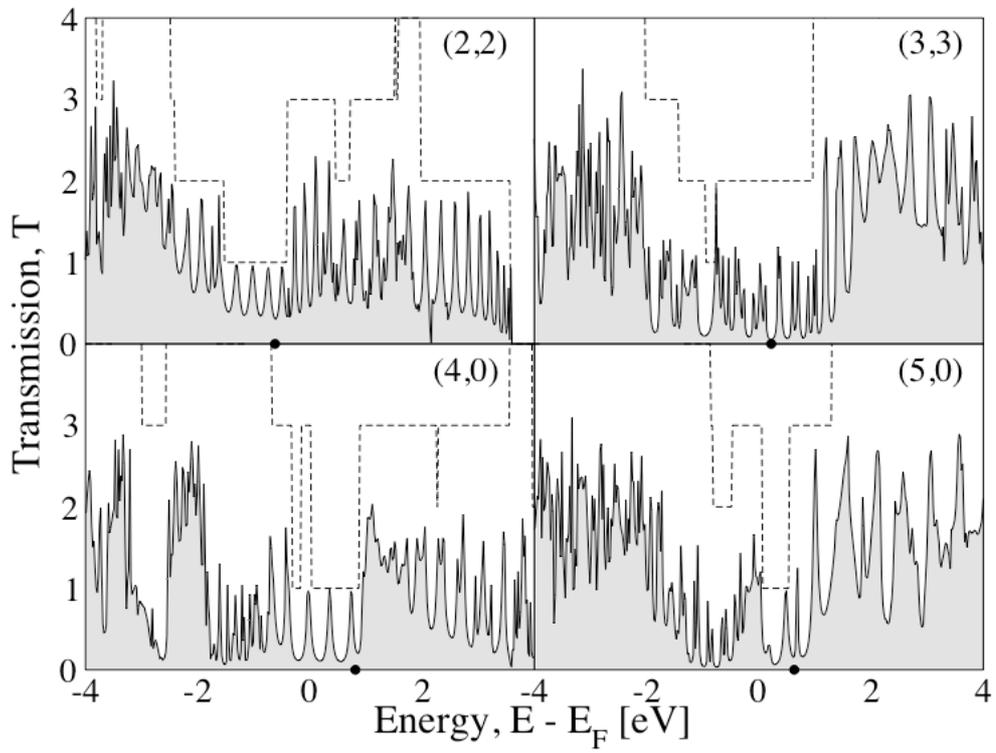

**Figure 5**

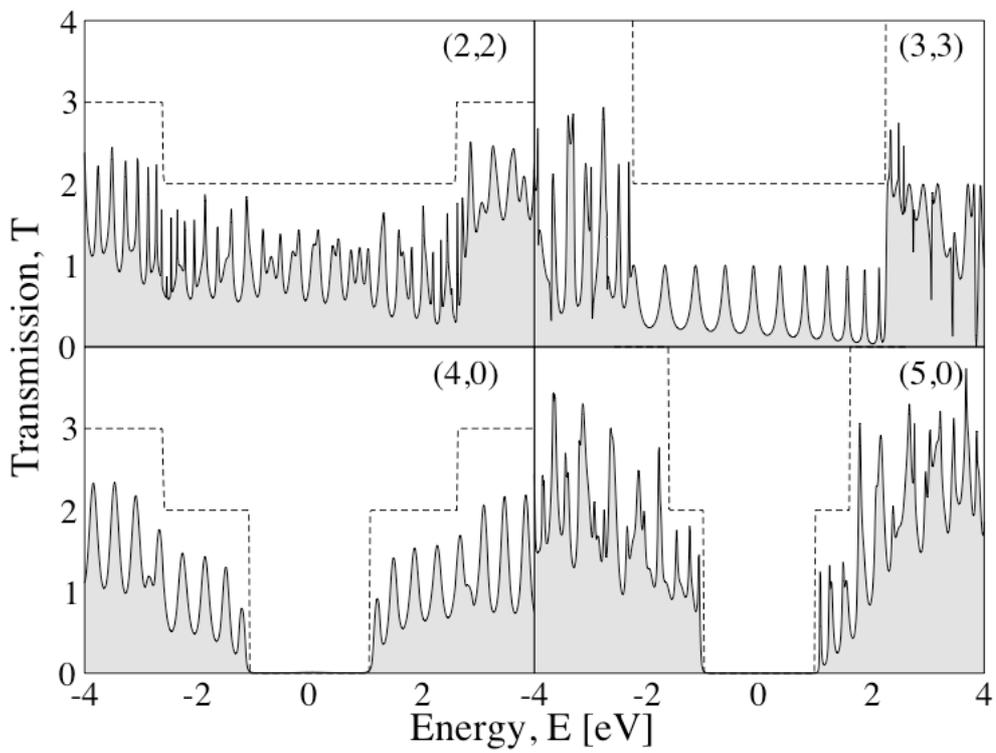

**Figure 6**

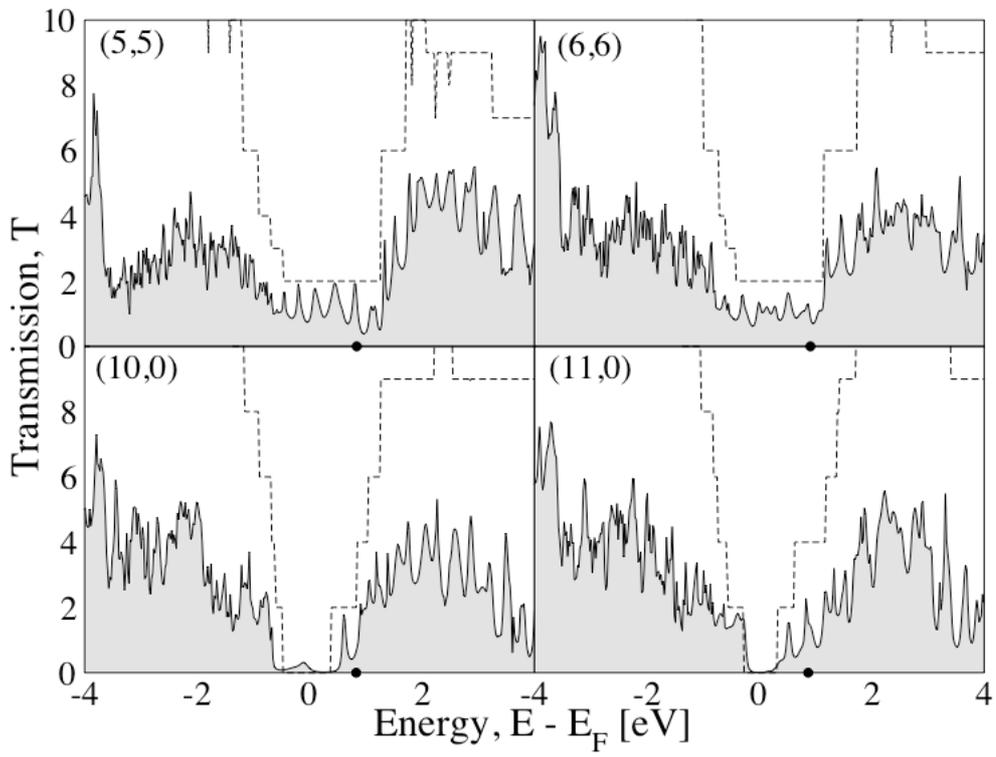

**Figure 7**

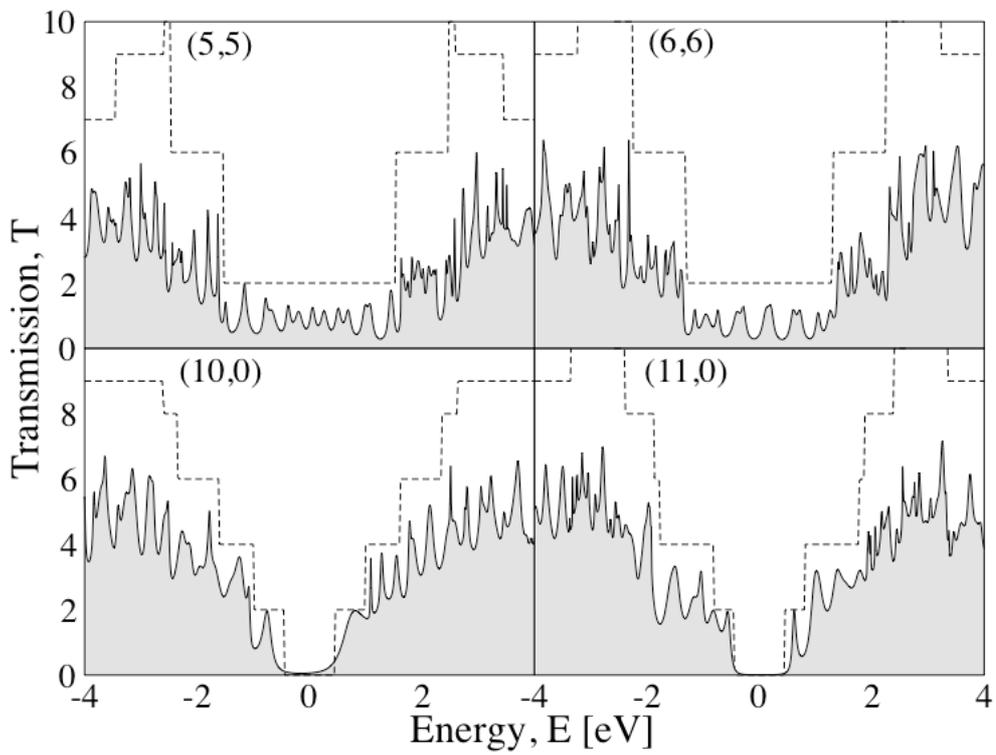

**Figure 8**

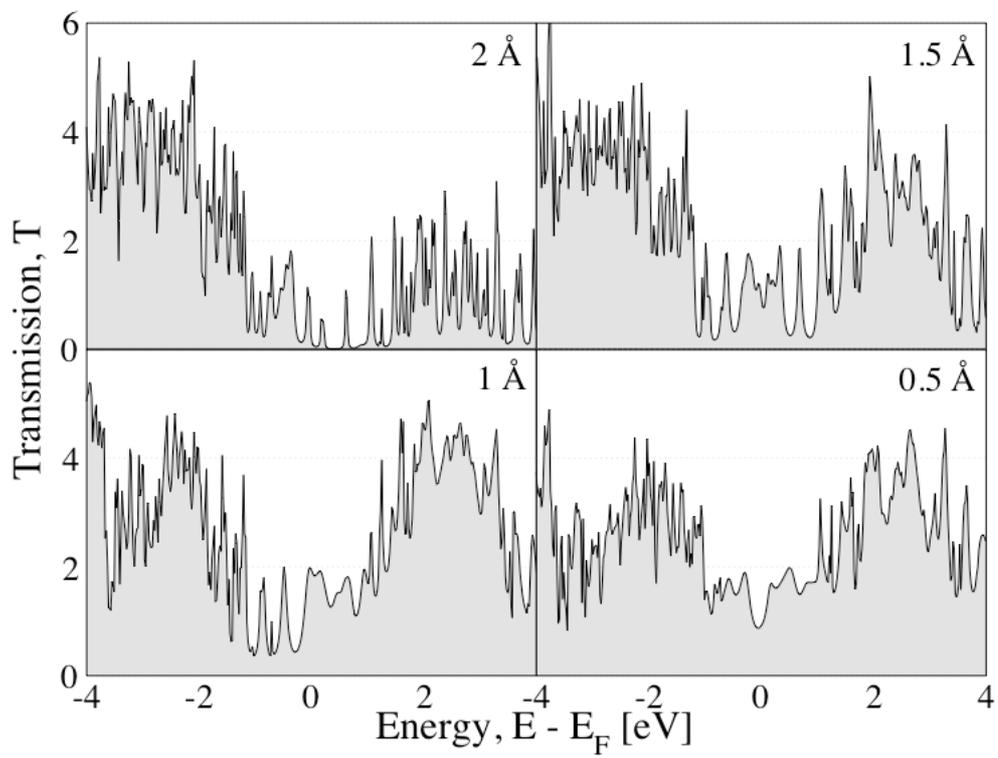

**Figure 9**

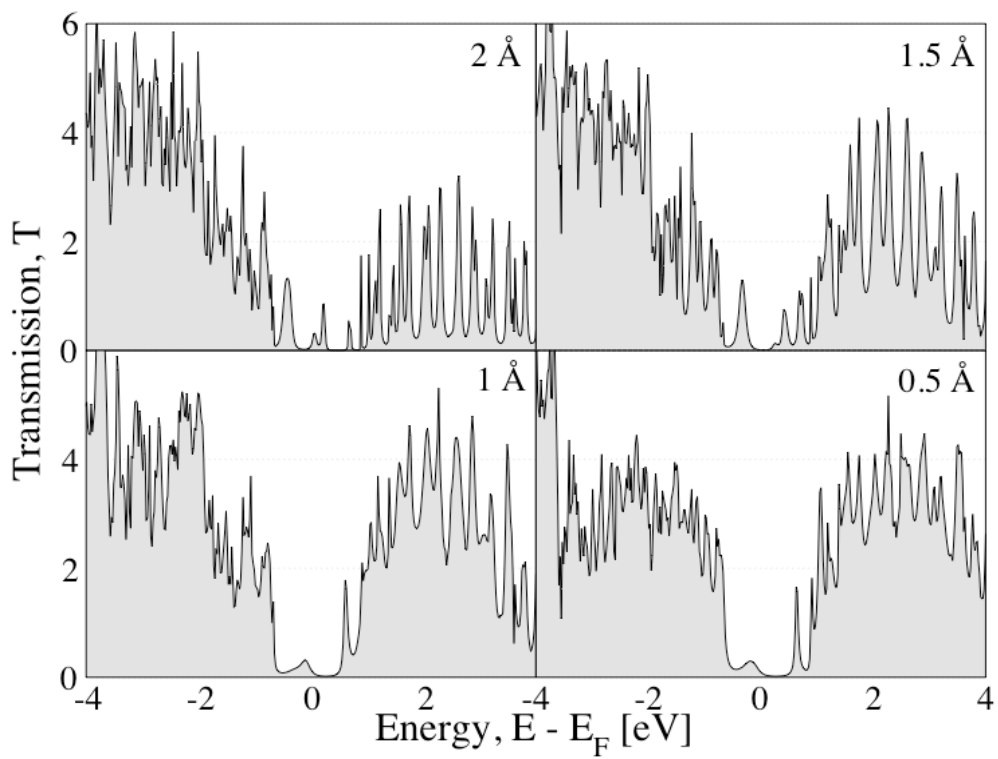

**Figure 10**

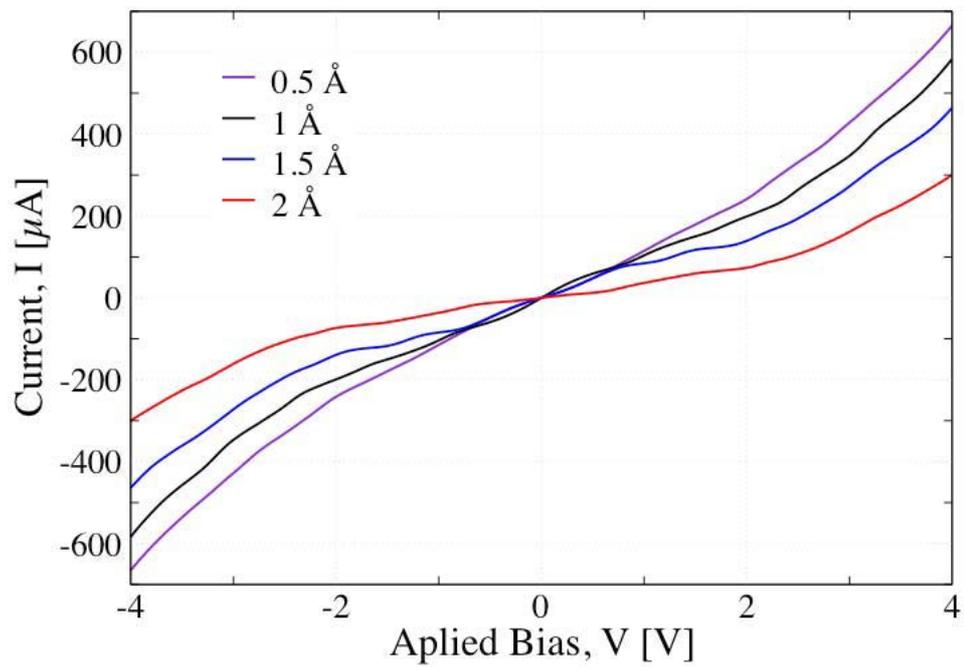

**Figure 11**

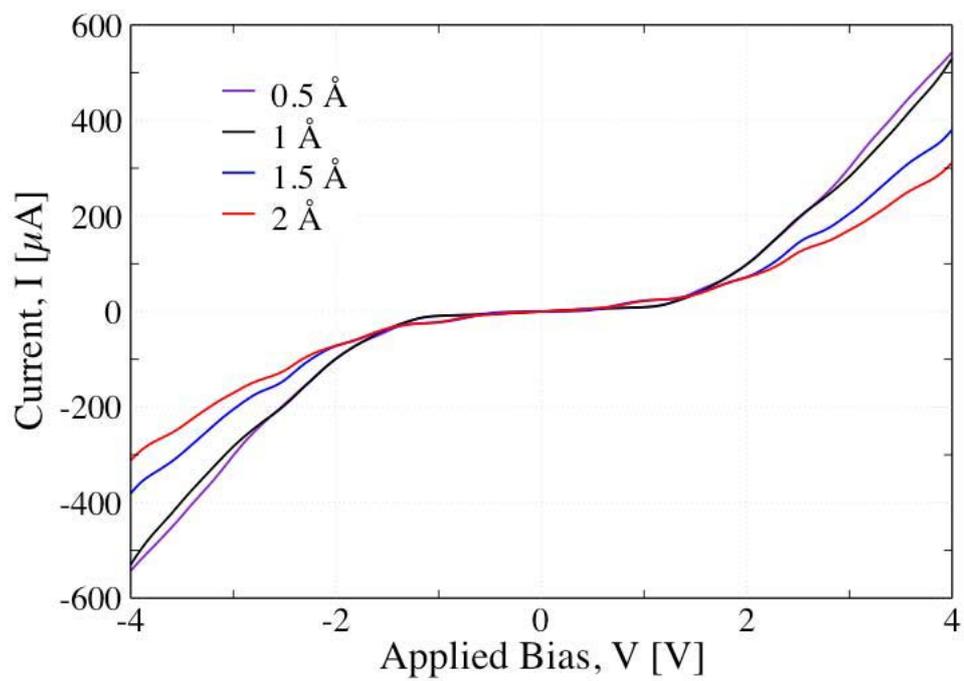

**Figure 12**

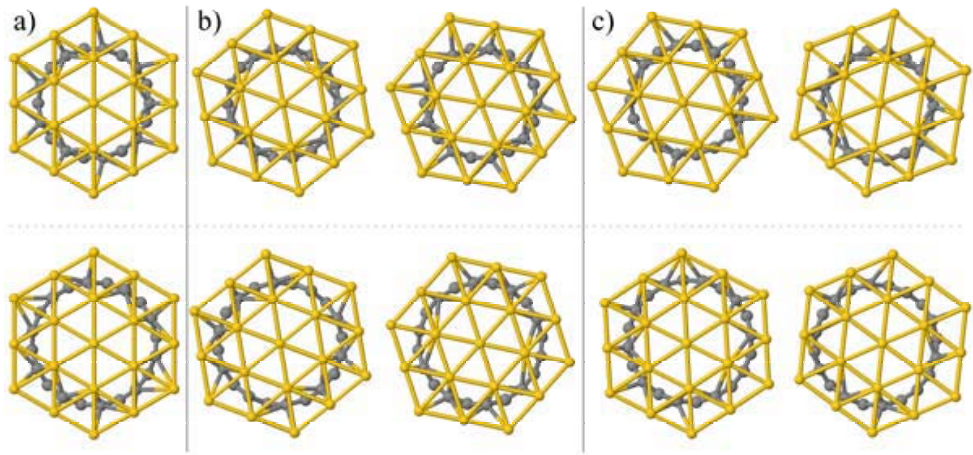

**Figure 13**

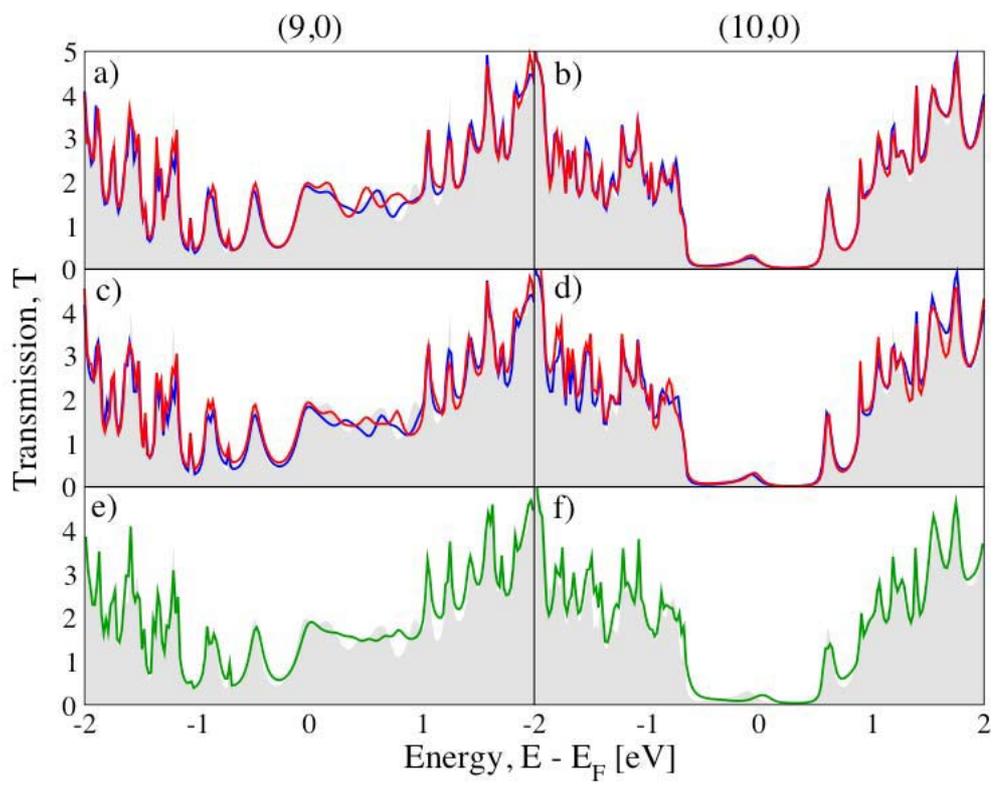

**Figure 14**

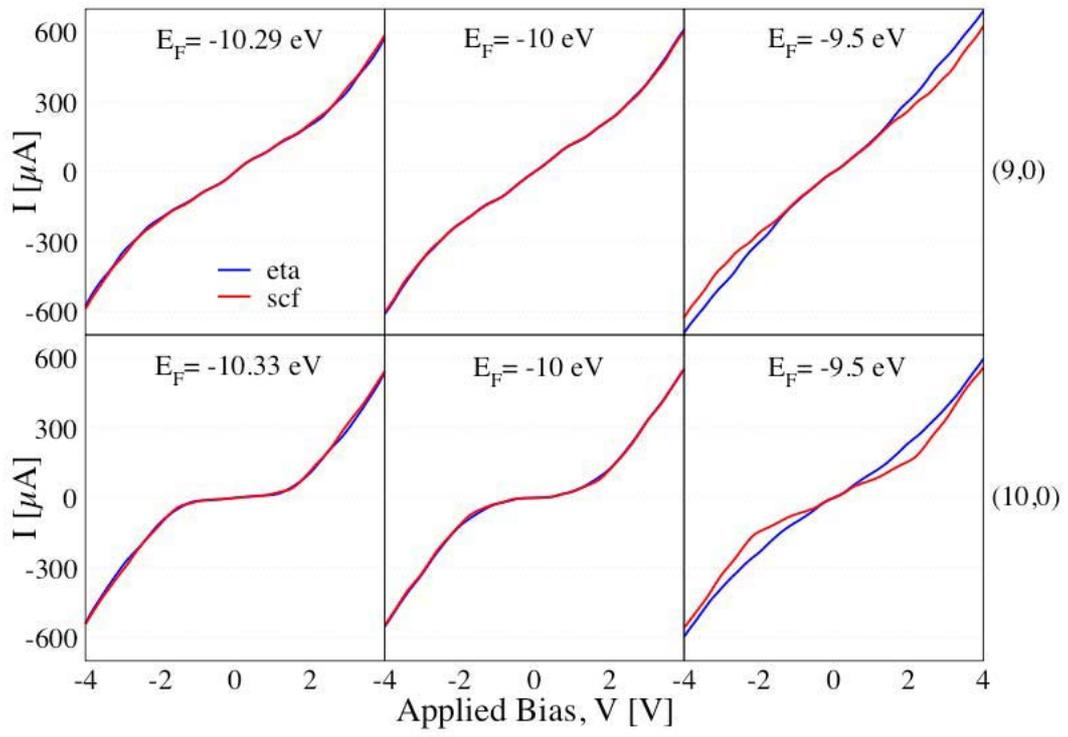

**Figure 15**